\documentclass{emulateapj} 
\slugcomment{Submitted to the Astrophysical Journal} 

\shorttitle{Simulations of Shock-Cloud Interactions} 
\shortauthors{Patnaude \& Fesen} 

\begin{document} 

\title{Model Simulations of a Shock-Cloud Interaction in the Cygnus Loop} 

\author{Daniel J.~Patnaude\altaffilmark{1,2} \& Robert A.~Fesen\altaffilmark{1}} 
\altaffiltext{1}{6127 Wilder Laboratory, Physics \& Astronomy
Department, Dartmouth College, Hanover, NH 03755}
\altaffiltext{2}{Smithsonian Astrophysical Observatory, 60 Garden St,
Cambridge, MA 02138}

\begin{abstract} 

We present optical observations and 2D hydrodynamic modeling of an isolated shocked
ISM cloud.  H$\alpha$ images taken in 1992.6 and 2003.7 of a small optical
emission cloud along the southwestern limb of the Cygnus Loop were used to measure
positional displacements of $\sim$ $0 \farcs 1$ yr$^{-1}$ for surrounding Balmer
dominated emission filaments and $0\farcs025 - \farcs055$ yr$^{-1}$ for internal cloud
emission features. These measurements imply transverse velocities of $\simeq$ 250 km
s$^{-1}$ and $\simeq$ 80 -- 140 km s$^{-1}$ 
for ambient ISM and internal cloud shocks
respectively.  A lack of observed turbulent gas stripping at the
cloud--ISM boundary in the  H$\alpha$ images suggests that there
is not an abrupt density change at the cloud--ISM boundary. Also, the complex
shock structure visible within the cloud indicates that the cloud's internal
density distribution is two phased: a smoothly varying background density
which is populated by higher density clumps. 

Guided by the H$\alpha$ images, we present model results for a shock
interacting with a non-uniform ISM cloud.  We find that this cloud can be 
well modeled by a smoothly varying power law core with a density contrast
of $\sim$ 4 times the ambient density,
surrounded by a low density envelope with a Lorentzian profile. 
The lack of sharp density gradients
in such a model inhibits the growth of Kelvin-Helmholtz instabilities, 
consistent with the cloud's appearance. 
Our model results also suggest that cloud clumps have densities
$\sim$ 10 times the ambient ISM density and account for
$\sim$ 30\% of the total cloud volume. Moreover, the observed spacing of internal
cloud shocks and model
simulations indicate that the distance between clumps is $\sim$ 4 clump
radii. We conclude that
this diffuse ISM cloud is best modeled by a smoothly varying, low density 
distribution coupled to higher density, moderately spaced internal clumps.

\end{abstract} 
\keywords{ISM: individual (Cygnus Loop) --- supernova remnants --- ISM: kinematics and dynamics --- shock waves --- hydrodynamics} 

\section{Introduction} 

The interaction of shock waves with the interstellar medium (ISM) such as those
associated with supernovae, stellar winds, bipolar flows, \ion{H}{2} regions,
or spiral density waves is a fundamental process in interstellar gas dynamics
and is key to understanding the evolution and structure of the ISM.
The highly nonlinear interaction between supernova generated shocks and 
interstellar clouds is often not suited to
analytic approaches but requires a multidimensional hydrodynamics study of the
shock-cloud problem using high resolution methods. 

A hydrodynamical study of a
shocked ISM cloud was made by \citet[][hereafter KMC94]{klein94}, who found
that the cloud may be destroyed by a series of instabilities associated with
the post-shock flow of inter-cloud gas past the cloud. Earlier work on this
problem includes that of \citet{woodward76} and \citet{nittman82}.  More
recently, \citet{poludnenko02} studied the role that internal cloud structure
plays in the destruction of the cloud.

For investigating shocked ISM cloud physics, the interaction between a supernova (SN)
shock and low density diffuse ISM clouds is of particular interest.  Supernova remnants
(SNRs) shape and enrich the chemical and dynamical structure of the ISM which,
in turn, affect the evolution of subsequent SNRs. The details of just how SN
generated shock waves interact with interstellar
clouds are not well understood.

There are several limiting factors in attempting to compare model
simulations to observed SNR shock cloud interactions. While models can be
viewed edge on and rotated in two or three dimensions, shocked interstellar
clouds are viewed only in projection, which leads to a complex appearing shock
structure due to multiple and overlapping shocks. In
addition, one observes only a single epoch, i.e., a `snapshot', of the
interaction. These factors make it difficult to understand and model the time
dependent kinematics and detailed dynamical processes of the interaction. Also,
unlike how they are often modeled, real interstellar clouds are neither
cylindrical or spherical in shape nor sharp edged, with interiors very
likely non-uniform in density.  Furthermore, many shocked interstellar clouds are
dense enough so that radiative losses, which can alter the overall 
dynamics of the shock-cloud interaction, are important \citep{mellema02,fragile04}.
Finally, the inclusion of an embedded magnetic field can drastically alter the
dynamics of the interaction. For instance, a strong, ordered magnetic field 
can suppress dynamical instability growth predicted by fluid dynamical simulations
\citep{maclow94,fragile05}.

In looking for an `ideal' shock-cloud interaction, one would like to avoid many
of the aforementioned effects and the Cygnus Loop supernova remnant affords
several distinct advantages. Because of the remnant's large angular size
(2.8$^{\circ} \times$ 3.5$^{\circ}$), low foreground extinction ($E[B-V] =
0.08$ mag; \citealt{parker67,fesen82}), and wide range of shock conditions, the
Cygnus Loop is one of the better locations for studying the ISM shock physics
of middle-aged remnants. At a distance of 550$_{-80}^{+110}$ pc
\citep{blair05}, it has a physical size of 27 $\times$ 33.5 pc. Located
8.5$^{\circ}$ below the galactic plane, the remnant lies in a multi-phase
medium containing large ISM clouds with a hydrogen density of $n=5-10$
cm$^{-3}$, surrounded by a lower density inter-cloud component of
$n\approx0.1-0.2$ cm$^{-3}$ \citep{denoyer75}.

Recently, \citet{patnaude02} studied a small, isolated cloud along the
southwest limb of the Cygnus Loop which met many of the desired cloud
properties for investigating shock-cloud interactions. This cloud is relatively
small \citep[$\sim$ 2$\arcmin$ in radius; $0.32$ pc at $550$
pc;][]{blair05}, exhibits a fairly uncomplicated, line-of-sight internal
structure, and lies isolated from other shocked ISM clouds. Moreover, the
shock-cloud interaction is dominated by non-radiative, or `Balmer-dominated'
filaments, indicating that the cloud-shock dynamics is not significantly
affected by post-shock radiative losses.

Here we present a new analysis of this small shocked cloud. Proper motion
measurements and inferred shock velocities of individual filaments in and
surrounding the cloud are presented.  These results were used to estimate the
initial conditions for hydrodynamical model simulations of a shock interaction
with an unmagnetized, lumpy cloud.  In $\S2$, these new observations are
presented as well as the technique used to measure the filament proper motions.
Model parameter estimates are then discussed in $\S3$. Our
hydrodynamical models are presented in $\S4$, where proper
motion and density estimates are implemented in the model initial conditions.
Model results are presented in $\S5$, and they are compared to
the southwest cloud in $\S6$ with our conclusions in
$\S7$.

\section{Observations} 
\label{sect:3observ} 

Narrow passband H$\alpha$ images of the southwest region of the Cygnus Loop
were obtained on 7 July 1992 and 29 August 2003 using the MDM 2.4 m Hiltner
telescope. For the July 1992 images, four 600 s H$\alpha$ filter (FWHM = 80
\AA) exposures were acquired with a Loral 2048 $\times$ 2048 front side
illuminated CCD yielding a spatial resolution of $ 0 \farcs 343$ pixel$^{-1}$.  
Details of the 1992 observations and subsequent data reduction can be found in
\citet{patnaude02}. 

Two 1000 s H$\alpha$ filter (FWHM = 15 \AA) exposures were
taken in August 2003 with a SITe 2048 $\times$ 2048 back side illuminated CCD
with a resolution of $ 0 \farcs 275$ pixel$^{-1}$. Using IRAF\footnote{IRAF is
distributed by the National Optical Astronomy Observatories (NOAO), which is
operated by the Association of Universities for Research in Astronomy, Inc.
(AURA) under cooperative agreement with the National Science Foundation.}, the
data were bias-subtracted, flat-fielded, and cosmic-ray hits were removed. The
resulting 2003 epoch image is shown in Figure~\ref{fig:regions}. Globally, the cloud's
morphology is nearly identical to that seen in the 1992 images \citep[Figs. 2
\& 3][]{patnaude02}, but close inspection between the two epoch images
showed measurable proper motions for both internal cloud structures and the
surrounding thin shock front filaments.

\section{Analysis and Results}
\label{sect:3mod_est}

Following the procedure described by \citet{thorstensen01}, the coordinate
systems of the two H$\alpha$ images were aligned using DAOPHOT in conjunction
with the USNO-A2.0 catalog. The datasets were then rebinned to an effective
image scale of $0 \farcs 1$ pix$^{-1}$. This rebinning introduced a small
global offset of $-0 \farcs 07$ pix$^{-1}$ between the two images,
uniform across the entire field of view. 

\subsection{Proper Motion Measurements}

Individual filament regions for the proper motion analysis were selected based
on their projection onto the plane of the sky, the complexity of the filament and 
surrounding regions, and the brightness of the filament feature. 
Based on these criteria, 14 filaments within the cloud, including both Balmer dominated and 
radiative filaments, and 21 regions from the surrounding Balmer 
dominated shock front were chosen (see Figure~\ref{fig:regions}).

One-dimensional intensity profiles were extracted for each region and the
pixel shift in each shock filament was computed using the IRAF task {\it xcsao},
which is based on the software of \citet{tonry79}. While this task was written
to compute relative radial velocities via the cross-correlation
function between two spectra, the cross-correlation function yields accurate 
filament motions in terms pixel shifts between two images. 
For thin Balmer-dominated filaments and bright and sharp cloud shock features,
the cross-correlation analysis was able to match the shock fronts between the
two epochs and measure the pixel shift between the two data sets to an accuracy
of 10\% -- 15\%. The results from this analysis are listed in terms of proper
motion (mas yr$^{-1}$) and transverse velocity (km s$^{-1}$) in
Table~\ref{tab:cc}. The quoted velocities assume a distance of
550$_{-80}^{+110}$ pc \citep{blair05}.

Example data and cross-correlation functions for two regions are shown in
Figure~\ref{fig:cc_sf14}. These one-dimensional filaments and cross-correlation
functions are representative of the data for the non-radiative filaments 
(Fig.~\ref{fig:cc_sf14}, left) and internal cloud filaments 
(Fig.~\ref{fig:cc_sf14}, right), where there is often a plateau of emission
(from shocked cloud material) downstream from the shock front and then 
a steep rise in emission at the cloud shock front.

Quoted errors in Table~\ref{tab:cc} include the signal to noise in the 
filament region, the curvature of the filament region, the profile of the 
filament when convolved with the image PSF (FWHM$_{1992}$ = $1 \farcs 0$; 
FWHM$_{2003}$ = $ 0 \farcs 7$), and the contribution from the background, local 
nebulosity, and adjacent filaments. For well resolved filaments, the cumulative 
effect of these errors is $\sim$ 0.6 pixel, or about 10 km s$^{-1}$.

\subsection{Estimate of Cloud Parameters}

As discussed in \citet{patnaude02}, this shock-cloud interaction is nearly tangent
along the line of sight to the southwestern limb of the southern region of the 
Cygnus Loop.  The
cloud is being run over by the remnant's shock front which is moving roughly
east-west. We have divided the cloud into two regions: the ISM shock front and
the cloud-shock region. Based on the filament velocities listed in
Table~\ref{tab:cc}, we estimated an interstellar shock velocity of 250 km
s$^{-1}$ associated with the Balmer-dominated filaments. The wide range of
Balmer filament velocities observed ($140 - 260$ km s$^{-1}$) may be due in
part to density fluctuations around the cloud and the fact that only one
component of the filament velocity is measured. That is, for filaments which
are highly curved, the space velocity of the shock might be 200 km s$^{-1}$,
but the local {\it measured} velocity might be in a direction other than
perpendicular to our line of sight. Furthermore, though filaments were chosen
based upon selective criteria, factors such as low signal to noise as well as
adjacent, overlapping filaments contributed in some cases to a poorer 
cross-correlation between the two images. Nonetheless, our estimated shock velocity
$\simeq$250 km s$^{-1}$ is consistent with the X-ray shock velocity
of $\sim$ 300 km s$^{-1}$ inferred from the ROSAT PSPC measurements
\citep{patnaude02}.

The shocked cloud can be further divided into regions where
the cloud-shock is interacting with the cloud, and where it is interacting with
cloud clumps or ``cloudlets''. In general, the inferred shock
velocities vary widely (65--140 km s$^{-1}$). This suggests that the density
structure of the cloud is fairly complex, as the cloud-shock appears to have
been slowed less in certain areas relative to others.

Based on these measurements, we adopt a shock velocity range inside the cloud of $60 - 100$ km
s$^{-1}$. These estimated cloud-shock velocities in turn imply a range of density
contrasts in the cloud. Assuming ram pressure equilibrium, ($\rho_a v^2_s
\approx \rho_c v^2_{cs}$) the density contrast between the cloud and the
ambient medium, $\chi$ $\equiv$ $\rho_c / \rho_a$, is $\sim$ 4--17, with higher
values representing areas populated with cloudlets, and lower values
representing regions of low density within the cloud.

The low density nature of this cloud permits us to view its internal shock
dynamics. We have used the structure and spacing of the internal
shock fronts to estimate the clumpiness of the cloud. The easiest place to do
this is at the western nose of the cloud (Regions C8--C10). Measurements
suggest that the post-shock spacing of clumps in the cloud is $\sim$
10$\arcsec$ -- 30$\arcsec$. The upper limits corresponds to shocks which are
more highly evolved, while the lower limit corresponds to `small' shocks. Based
on the size of the cloud (2$\arcmin$ -- 4$\arcmin$ diameter), the cloudlets
likely account for 30\% of the total cloud volume. Furthermore, based on a
maximum compression of $\sim$ 4 for cloud clumps, we estimate the spacing 
between cloudlets to be $\approx$ 4--5 cloudlet radii ($a_{cloudlet}$).

\section{Hydrodynamical Models}
\label{sect:3model}

For modeling this shock-cloud interaction, we have
used the numerical hydrodynamics software VH-1 \citep{blondin92,stevens92},
which implements the 
piecewise parabolic method (PPM) to solve the equations of gas dynamics 
\citep{colella84}. The PPM approach incorporates a fixed computational grid to 
evolve the standard conservation equations of mass, momentum, and energy. We 
assume an ideal, inviscid fluid with a ratio of specific heats, $\gamma$, equal 
to $5/3$. VH-1 does not explicitly treat the collisionless shock physics
associated with Balmer-dominated shocks. However, the goal of this study is to 
understand how a blast wave interacts with a diffuse ISM cloud. For these purposes,
VH-1 serves as an excellent tool for tracing the motion and dynamics of this 
interaction.

Simulations were performed on a 1440 $\times$ 1440 Cartesian grid.
The fiducial physical size of the square grid is 1 $\times$ 1, with dx = dy = 
6.9 $\times$ 10$^{-4}$, The size of the cloud sets the scale of the models. 
The cloud is $\approx$ 2$\arcmin$ in radius ($\approx$
10$^{18}$ cm $\times$ d$_{550\,\,{\mathrm pc}}$). On average, the cloud radius 
is 30\% of the
computational grid, or $\approx$ 450 cells per cloud radius. Therefore, the
physical length scale of the grid $\Delta$x $\approx$ 2 $\times$ 10$^{15}$ cm
cell$^{-1}$. 

We estimate the importance of radiative losses by calculating the cooling 
time scale and comparing it to dynamically relevant time scales (mainly, the 
cloud crushing time and the pressure variation time scale). In general, 
radiative losses will be considered important if the cooling time is 
comparable to or shorter than the cloud crushing time. We estimate the 
cooling time using the approximation of \citet{kahn76}, $t_{cool}$ = 
$C v^3 / \rho$, where $v$ is the cloud shock velocity in units of km 
s$^{-1}$, $\rho$ is the cloud density in units of gm cm$^{-3}$, and $C$ is a 
constant $=$ 6.0 $\times$ 10$^{-35}$. Assuming a cloud shock velocity of 140 
km s$^{-1}$ and a cloud density of 10$^{-24}$ gm cm$^{-3}$, we estimate a 
cooling time $t_{cool}$ $>$ 5000 yr. In contrast, the cloud crushing time, 
$t_{cc}$ $\equiv$ $\chi^{1/2} a_0/v_b$, is $\approx$ 3800 yr, for a blast 
wave velocity of 250 km s$^{-1}$ and an initial cloud radius of $a_0$ = 
0.3 pc. Furthermore, the pressure variation time scale is $\sim$ 0.1$t_{cc}$ 
(KMC94), which is $<$ $t_{cool}$. Thus, it appears that neglecting the 
effects of cooling in our models will not have a significant impact on our 
results. This is supported by \citet{patnaude02} who showed that this cloud 
is only weakly radiative.

Under the assumptions that radiative losses are not dynamically important and that
magnetic fields are not present (or that the cloud is only weakly magnetized), the
shock-cloud interaction can be wholly defined by two parameters (KMC94), the 
shock Mach number, and $\chi$, the density contrast of the cloud.
Assuming an ISM sound speed of $\approx$ 10--15 km s$^{-1}$ and a blast wave 
velocity of 250 km s$^{-1}$, we estimate a 
shock Mach number of $M$ $\approx$ 20. Furthermore, as pointed out in $\S3$, we
estimate a cloud to ISM density contrast of $\chi$ $\approx$ 6. 

To further simplify the problem, 
we chose a set of non-dimensional variables such that the ISM density $\rho_a$ 
is set to the ratio of specific heats, $\gamma$ = $5/3$, and the ISM pressure, 
$P_a$, is set to unity. The ISM sound
speed, $c_a$, is thus set to 1 ($c_a$ = $(\gamma P_a/\rho_a)^{1/2}$), and the 
shock velocity $v_b$ is just the shock Mach number.

Model results by KMC94 suggested that a cloud radius should be at least of
order 120 cells.  For our models here, we chose the main cloud to have a radius
of 300-500 cells ($\simeq$ 10$^{17-18}$ cm). The internal cloudlets have 
sizes which
are 10--20\% of the cloud radius. Therefore, the cloudlets are only 33\%
the suggested size. This small cloudlet size limits our ability to resolve
instability growth along their boundaries. However, the goal here is to understand
the global, internal morphology of the cloud, rather than small scale mixing
within the cloud.

We broke the cloud's density structure into two parts: A background density 
profile, and a clumping or density perturbation distribution. The background density 
distribution is chiefly responsible for the large-scale shock features,
such as how the shock drapes over the 
cloud edges, while also defining the initial internal cloud shock velocity. 
In contrast, the internal cloud density
perturbations lead to the formation of small scale shock structures
within the cloud and have little effect on the cloud--ambient medium
boundary layer.

Based on the cloud's emission features (Fig.~\ref{fig:regions}), we assume that the 
large scale 
density structure of the cloud is smoothly varying. The interface between the blast wave
and the cloud shock, seen along the northern and southern periphery of the cloud,
appears smooth. This suggests that the cloud is surrounded by a low density envelope. 
There is no evidence to suggest that the central density is sharply peaked, so we assume
that at some inner radius the density profile turns over and becomes relatively constant
throughout. Therefore, we assume that the cloud consists of a smoothly varying core
surrounded by a low density envelope. A
function which fits this description is a truncated Lorentzian coupled to a
power law core:

\begin{equation}
\chi(r) = \cases{
\frac{\chi_{max}}{1 + r^2_i/r^2_0}\left(1 + 
\Delta\left(\frac{r_i - r}{r_i}\right)^{\alpha}\right) & $0 \leq r \leq r_i$ \cr
\frac{\chi_{max}}{1 + r^2/r^2_0} & $r_i < r \leq a_0$ \cr}
\label{eq:densdist}
\end{equation}

\noindent where $r^2_0 = a^2_0/(\chi_{max} - 1)$, $\Delta$ $=$ 0.11 sets the maximum
core density, and $\alpha$ is the power law index $=$ 0.5 which ensures 
continuity across the core-envelope interface. 

While the fine-scale structure of density perturbations within the cloud is not known, 
the lack of observed dynamical instabilities (at the resolution of the observations),
suggests that such perturbations are probably smoothly varying. Therefore,
for models using the above cloud density distribution, we chose to model the
cloudlets as Gaussians. The spacing of the Gaussians is such that $\Delta r$
$\approx$ 4$\sigma$ between the cloudlet cores, which is consistent with the
optimal spacing of $\Delta r$ $\approx$ $4.2a_{cloudlet}$ suggested by
\citet{poludnenko02}. Individual model parameters are
listed in Table~\ref{tab:modparms}. For comparison, we include models of cylindrical
clouds with similar density distributions.

\section{Results}
\label{sect:3results}

Our basic shock-cloud interaction, Model 1, is shown in the top panel of
Figure~\ref{fig:r1d34}.
This model is of a Mach 20 shock interacting with
a cloud of radius $a_0$ = 0.3 and constant density contrast $\chi$ = 6.
Figure~\ref{fig:r1d34} shows the model 
at $t$ = 1.4, 2.6, 3.6, and 4.7 $\times$ 10$^{-2}$. Panel {\it c} ($t = 3.6 
\times 10^{-2}$), shows the model at approximately one cloud 
crushing time (the cloud crushing time, given by Equation 2.3 of KMC94,
is $t_{cc}$ $\approx$ 3.7 $\times$ 10$^{-2}$). According to 
KMC94, the growth time for Kelvin-Helmholtz (K-H) instabilities 
is $t_{KH}$ $\approx$ $\chi^{1/2} /(k v_{rel})$, where $v_{rel}$ is the 
relative velocity between the shocked cloud and the shocked ambient medium. 

The relative velocity between the post cloud-shock material and the post 
shock ambient material is given to first order by the relative jump 
conditions between the cloud and the ambient medium. Since $M$ $\propto$ 
$\chi^{1/2}$, $t_{KH} \sim \chi^{1/2} t_{cc}$ (for $k \sim a_0$). Higher 
wavenumber perturbations will form on a shorter time-scale. In Panel {\it c} 
of Figure~\ref{fig:r1d34}, there is clear evidence for K-H growth along the 
backside of the cloud. In fact, it is evident that K-H growth occurs much 
earlier (top, Panel {\it b}, Fig.~\ref{fig:r1d34}). 

Model 2 is shown in the bottom panel of 
Figure~\ref{fig:r1d34}. This model has the density
distribution described in Equation~\ref{eq:densdist}, with an inner 
radius $r_i$ = $0.6a_0$. The evolution of this model is markedly different 
than that of Model 1 (Fig~\ref{fig:r1d34}). The $\chi$ = 10 listed in 
Table~\ref{tab:modparms} is what the density would be at the center of
the Lorentzian. However, since the Lorentzian is truncated at 
$r_i$, the effective $\chi$ is much lower, by an amount $1/(1 + r^2_i/r^2_0)$.
Therefore, the maximum $\chi$ in the cloud is $\approx$ 3, or half the
$\chi$ of Model 1.
More importantly than the lower $\chi$ between the two models, Model 2 does 
not show signs of the instability growth seen in Model 1. This
interesting result lends weight to the notion of a smooth boundary between
the ISM and an embedded ISM cloud.

While Models 1 and 2 accentuate the differences which arise between a 
smoothly varying density distribution and that of a sharp edged cylinder,
the remaining models (3--6) simulate how the internal density structure 
affects the cloud shock.
Models 3 and 4 (Fig.~\ref{fig:r6r7}) are of cylindrical
clouds of $\chi$ = 6 and 3. Both clouds contain perturbations with 
a $\chi$ of 10 above the cloud density (or 30 and 60 times the ambient 
density). In Model 3 ($\chi$ = 6), the flow around the cloud is still strongly
influenced by the higher cloud density. The inclusion of cloudlets results
in the formation of shock structure within the cloud (not seen in 
Model 1). However, in the late time (Panel d), the morphology of the cloud
is still similar to the late time morphology of Model 1. 

The evolution of Model 4, on the other hand, is more strongly influenced by
the cloudlets. This is because the density contrast between the cloud and 
the ambient medium here is only 3, and thus the cloud shock velocity is not 
significantly different than the blast wave velocity ($\sqrt{3}$ lower). 
Instead, the blast wave is more influenced by the high density cloudlets (relative
to the cloud),
as seen in Figure~\ref{fig:r6r7} ({\it bottom}). While this model reproduces the 
observed shock diffraction (c.f. Fig.~\ref{fig:regions}, the presence
of instability growth along the (albeit low density) model cloud boundary
is not something observed in the observations.

Models 5 and 6 represent our best approximations to this diffuse ISM cloud. 
Physically, the model distributions represent cool, low density clouds 
surrounded by warm, lower density envelopes. Within the clouds, cold 
dense cloudlets are interspersed here on a regular grid. Cloudlet
formation is beyond the scope of this paper but is likely a thermal,
rather than gravitational condensation.

Model 5, shown in Figure~\ref{fig:d40d41} ({\it top}), is of a Mach 10 
shock over-running
a cloud with a density distribution given by Equation~\ref{eq:densdist}. 
The cloudlets have a $\chi$ of 10 and a maximum extent of $a_{cloudlet}$ = 
$0.05$. 
Furthermore, the inner radius of the cloud core is $0.6a_0$, which results in some of
the cloudlets being outside of the cloud core. 

As seen in Figure~\ref{fig:d40d41}, the blast wave shock is hardly slowed
by it's interaction with the cloud, similar to Model 4. 
However, the high density cloudlets
do significantly alter the cloud shock structure. Compared to Model 2 
(Fig.~\ref{fig:r1d34}), the cloudlets appear to play a significant role in 
slowing the cloud shock.

Model 6 differs from Model 5 in three ways: First, the $\chi$ of the cloud 
is 8, rather than 10; secondly, the $\chi$ of the cloudlets is increased
to 15, and thirdly, the radius of the cloudlets is $0.03$. Model 6 is shown
in Figure~\ref{fig:d40d41} ({\it bottom}). 
Here one sees that the $\chi$ of the cloud is so low
that it barely slows the shock. Moreover, and probably more importantly,
the spacing of the cloudlets is
such that they do not feel the effects of their neighbors (i.e. $\Delta r$
$>$ $4.2a_{cloudlet}$ and the diffracted shocks are not significantly curved).

Several features appear in the model simulations which are not observed. Prominent
in all the models is the formation of a bow shock behind the cloud. A bow shock
is not seen in Figure~\ref{fig:regions} simply because it is moving back into 
previously ionized material, so that there is no neutral population to excite. In
models containing cloud clumps (Models 3--6), fingers and mushroom heads are prominent
in the post shock flow. Three-dimensional models for shock cloud interactions show
that many of these features are unstable and will not persist in three dimensions due
to turbulent effects in the post shock flow \citep{stone92}.

\section{Discussion}
\label{sect:3disc}

As seen in Figure~\ref{fig:regions}, the southwest cloud of the Cygnus Loop represents 
a fairly uncomplicated case for investigating 
many of the basic phenomena of a shock-cloud interaction. The low density of 
the cloud implies that the cloud shocks will be largely non-radiative in 
nature. Compared to other regions of the Cygnus Loop 
\citep[and references therein]{levenson01}, the low density nature of this 
small cloud allows us to view internal cloud shocks. Furthermore, while 
the east-west extent of the cloud is not known, the location of the forward 
shock not interacting with the cloud suggests that this shock-cloud interaction 
is relatively young. Thus, this cloud presents a good test case to 
model the interaction between a SNR shock and an ISM cloud.

\subsection{Comparisons to Other Shock Models}

There have been several previous studies on shock-cloud interactions. Perhaps the 
best current model for comparison is that of KMC94. Though there is not a 
one-to-one comparison between our Model 1 and their models due to the differing 
initial condition, many of their conclusions are observed in Model 
1 such as the formation of K-H instabilities on the order of $t_{cc}$ 
({\it top}, Panel {\it c}, Figure~\ref{fig:r1d34}), like that found in KMC94.

On the other hand, there have been few published studies concerning the 
interaction between a shock and a cloud with a smoothly varying density. Our 
models seen in Models 2, 5, and 6 suggest that much of the instability growth 
observed in previous studies is related to the chosen geometry. ISM clouds are 
often modeled as cylinders or spheres with sharp, well defined boundaries. 
Yet, the real 
boundary between the ISM and embedded diffuse clouds is likely to be less 
distinct. However, models where 
the density varies over a certain distance such as that described by
a hyperbolic tangent \citep{poludnenko02} can sometimes lead to 
spurious instability 
growth like that seen in the sharp edged cylindrical case. 

In regard to the internal cloud density structure (`cloudlets'), 
\citet{poludnenko02} found that the principle parameter is the spacing 
between the cloudlets. Their models suggested that there exists a critical 
separation between cloudlets of $\approx$ 4.2$a_{cloudlet}$, and not 
surprisingly, the cloudlet spacing in Model 5 is about this value. 
Furthermore, as pointed out by \citet{poludnenko02}, a larger 
$\chi_{cloudlet}$ combined with a larger cloudlet spacing does not result in 
dynamics which are similar to the case of a lower $\chi_{cloudlet}$ combined 
with a smaller cloudlet spacing. Instead, as evidenced by Model 6, the larger 
separation, regardless of $\chi_{cloudlet}$, results in what are essentially 
multiple, independent interactions between the cloudlets and the cloud-shock.

\subsection{The Southwest Cloud}

\begin{figure}
\plotone{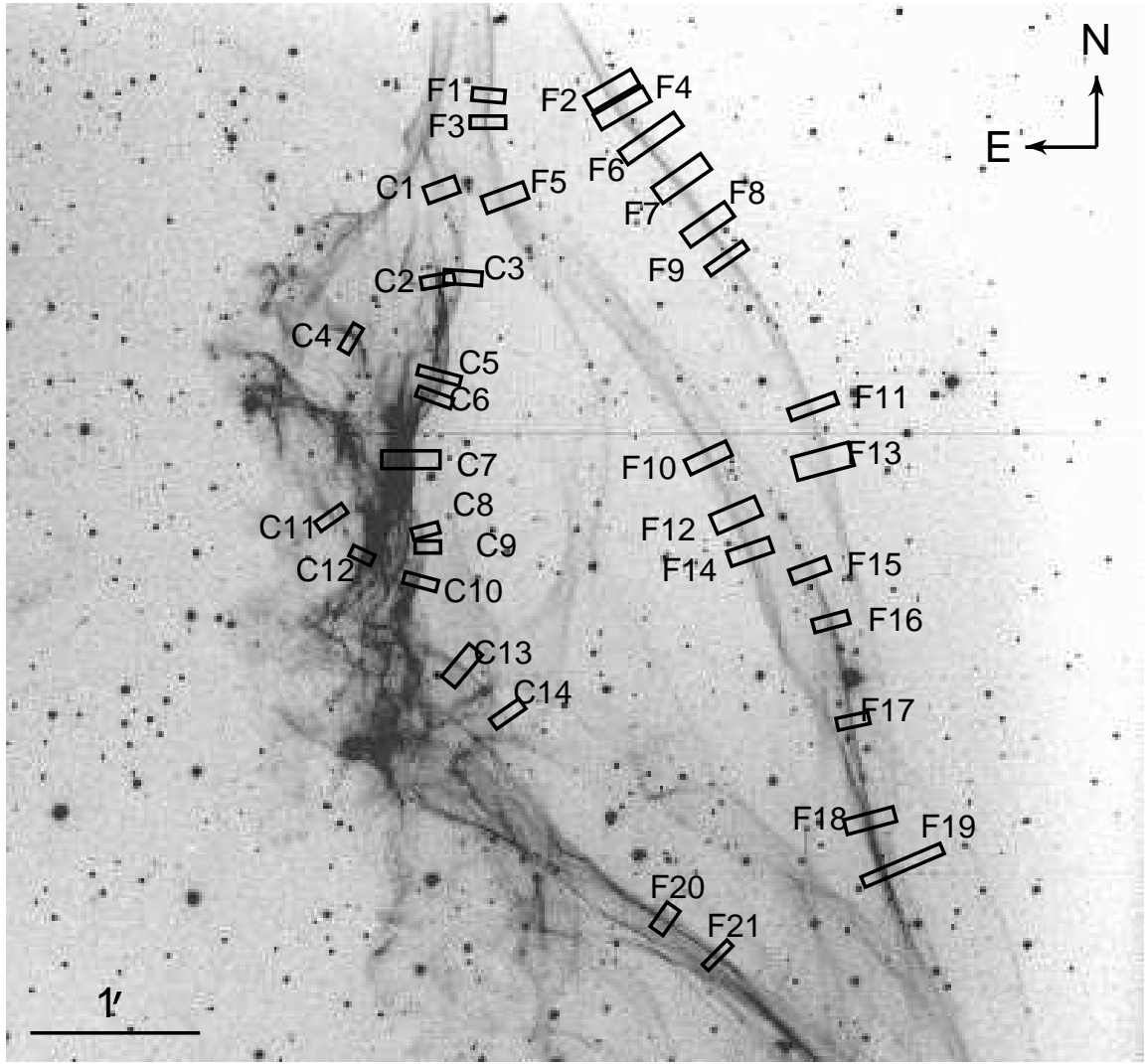}
\caption{August 2003 MDM 2.4m H$\alpha$ image of the southwest cloud in the 
Cygnus Loop showing the location of our selected filaments.}
\label{fig:regions}
\end{figure}

\begin{figure}
\plottwo{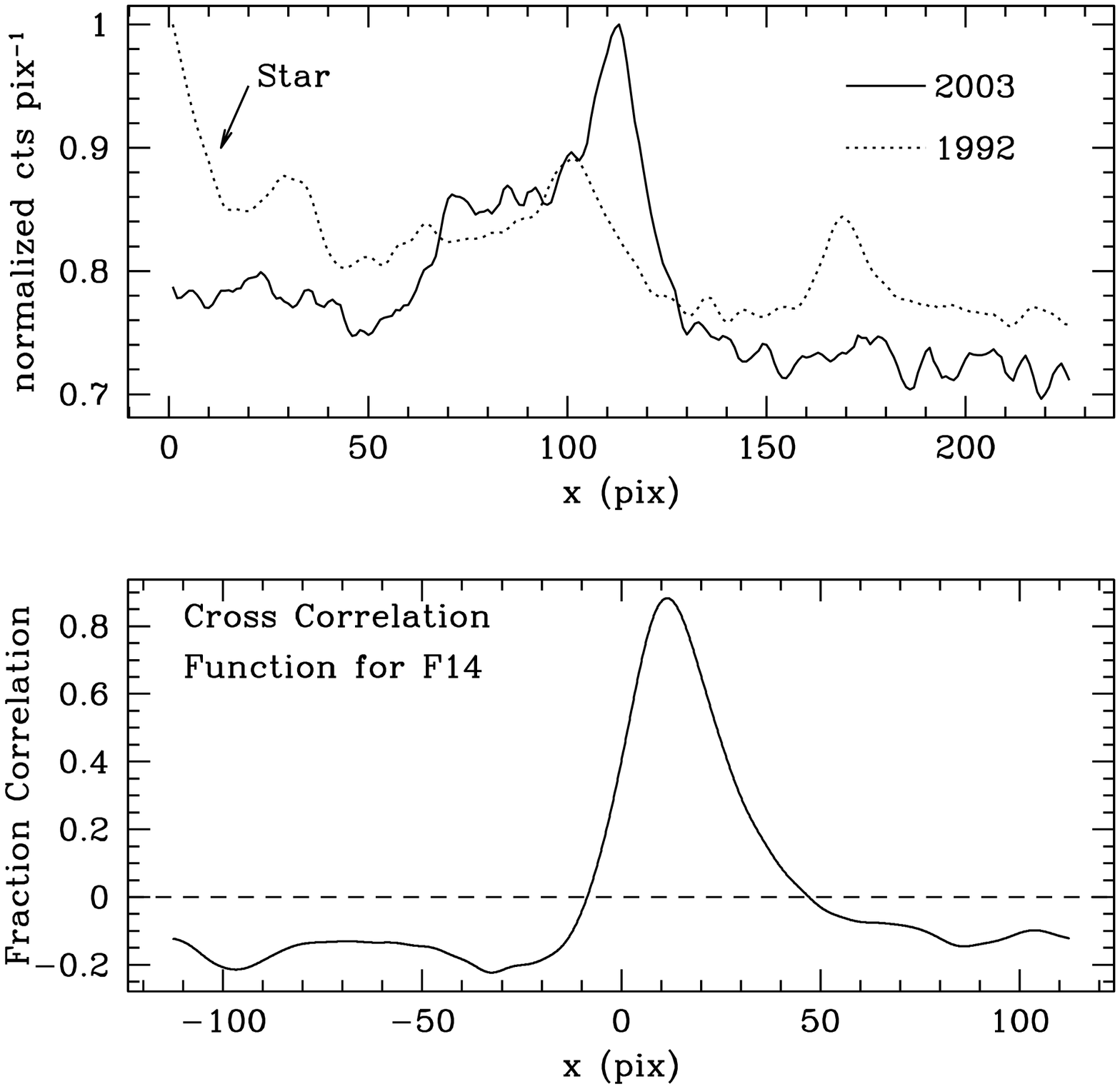}{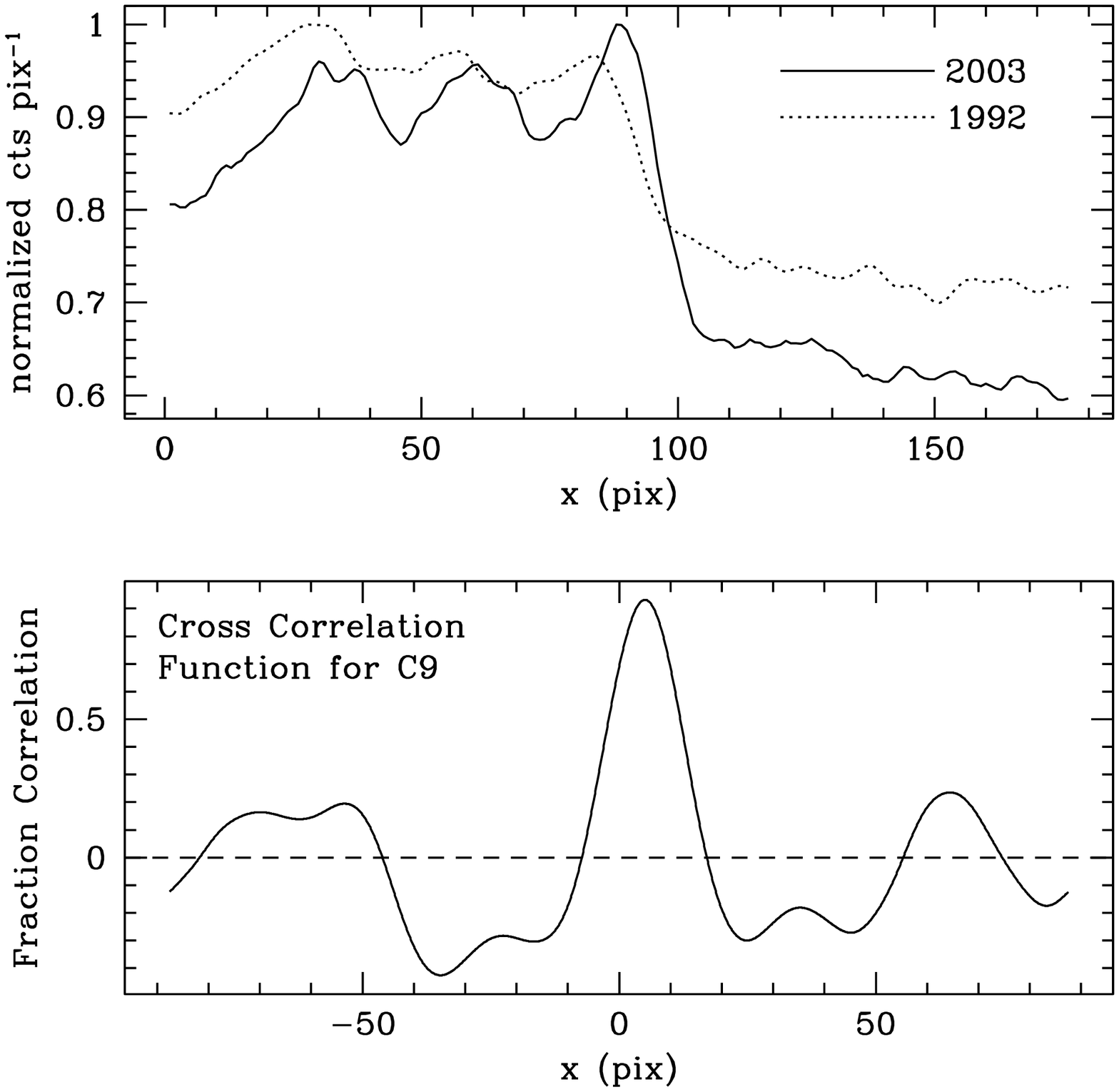}
\caption{{\it Left}: One-dimensional shock profiles ({\it top}) and 
cross-correlation function({\it bottom}) for region F14. The star appearing in 
the 1992 data 
(top, extreme left) is the result of using a wider filter during the 1992 
observations. {\it Right}: Cloud-shock profile ({\it top}) and 
cross-correlation function ({\it bottom}) for region C9.}
\label{fig:cc_sf14}
\end{figure}

\begin{figure}
\includegraphics[
angle=270,scale=0.6]{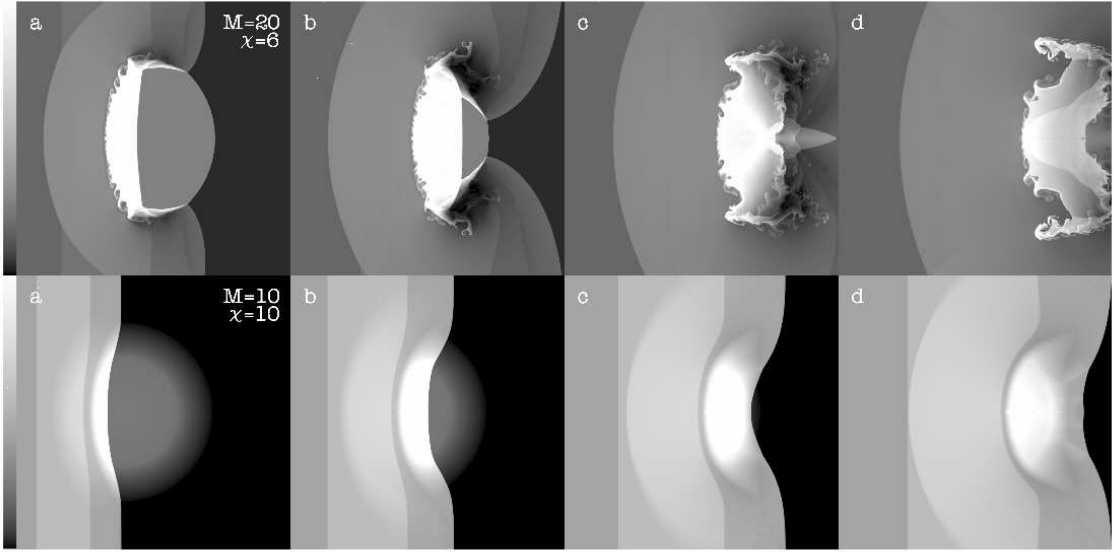}
\caption{Density plots of Model 1 ({\it top}) and Model 2 ({\it bottom)}. Model 
1 is shown at $t$ = 1.4, 2.6, 3.6, and 4.6 $\times$ 10$^{-2}$. Model 2 is 
shown at $t$ = 3.7, 5,5, 6.9 and 8.2 $\times$ 10$^{-2}$.}
\label{fig:r1d34}
\end{figure}

\begin{figure}
\includegraphics[
angle=270,scale=0.6]{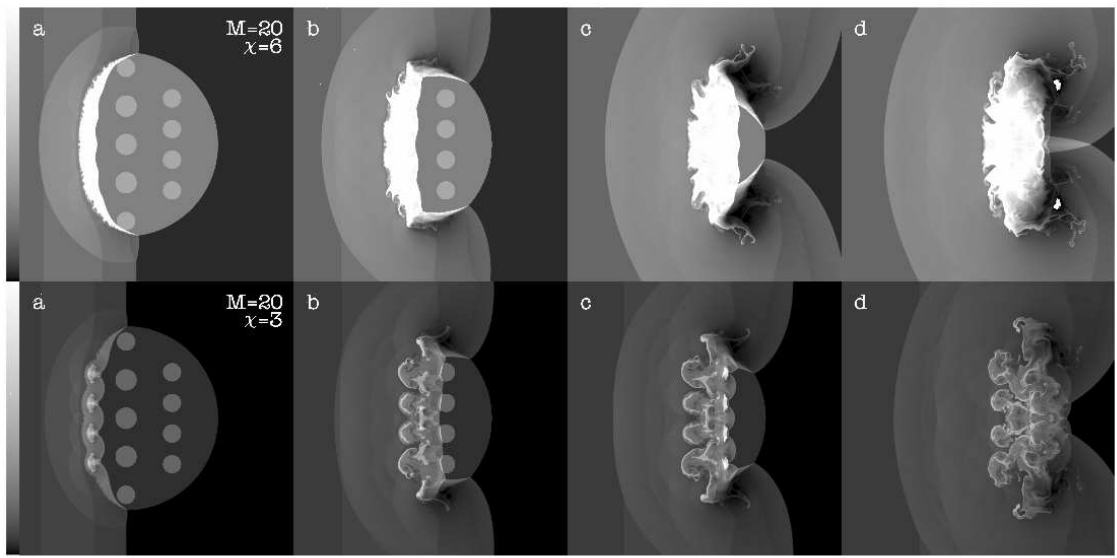}
\caption{Same as Figure~\ref{fig:r1d34} for Model 3 ({\it top}) and 
Model 4 ({\it bottom}). Model 3 is shown 
at 1.8, 2.9, 4.0, and 4.7 $\times$ 10$^{-2}$. Model 4 is shown
at 1.7, 3.0, 3.3, and 4.3 $\times$ 10$^{-2}$.}
\label{fig:r6r7}
\end{figure}

While the southwest cloud represents a valuable laboratory for 
investigating the shock-cloud interaction, as evidenced in 
Figure~\ref{fig:regions}, it is still highly complex on 
small scales. Hence, the models presented here only approximate its global 
properties. Based on Figure~\ref{fig:regions}, the cloud has a radius 
(N-S direction) of 1--2$\arcmin$. At the assumed distance of 550 pc, this 
corresponds to 0.16--0.33 pc, or $\approx$ 0.5--1 $\times$ 10$^{18}$ cm. 
In Model 5, the fiducial radius of the cloud is $0.35$. Using Model 5 as our 
potential model for the southwest cloud, the 
length scale of the model is thus 1.5 $\times$ 10$^{15}$ cm cell$^{-1}$.

The density of the ISM in this region is estimated to be $\sim$ 0.1 -- 0.3 
cm$^{-3}$. The maximum $\chi$ for Model 5 is 10, but in 
reality the density profile is truncated at an inner radius $r_i$ = 0.21$a_0$.
At $r_i$ = 0.21$a_0$, $\chi$ $\approx$ 4 for Model 5. This
agrees with our lower density estimate of $\chi$ $\approx$ 4.5 from ram pressure
arguments. Thus, the cloud particle density is $\approx$ 0.4 -- 1.2 cm$^{-3}$ with a
$\chi$ = 10 for the individual clumps in Model 5 ($n$ $\approx$ 1.0 -- 3.0 cm$^{-3}$).
The lower shock velocities seen in the cloud suggest cloudlet $\chi$'s as 
high as 17, but the difference between a Gaussian profile with a peak density
of 10 and one of 17 is minimal. 

\begin{figure}
\includegraphics[
angle=270,scale=0.6]{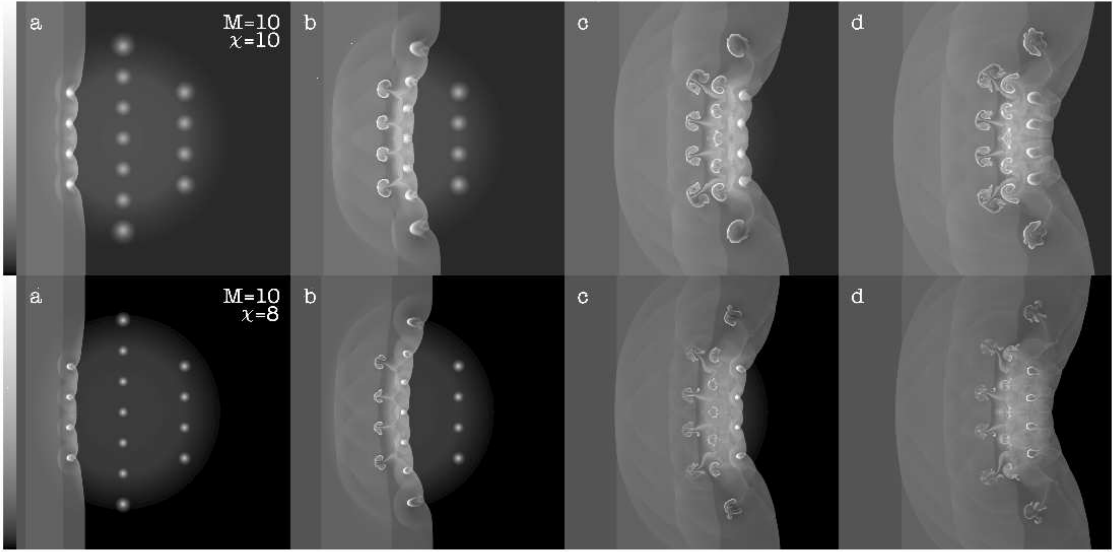}
\caption{Same as Figure~\ref{fig:r1d34} for Model 5 ({\it top}) and 
Model 6 ({\it bottom}). Model 5 is shown 
at 2.3, 4.6, 6.6, and 7.5 $\times$ 10$^{-2}$. Model 6 is shown at 2.3, 
4.3, 6.4, and 7.4 $\times$ 10$^{-2}$.}
\label{fig:d40d41}
\end{figure}

Based on the size of the grid cell and the shock velocity, the ambient shock 
traverses one cell in $\sim$ 10$^{8}$ s $\simeq$ 3 yr. The time difference in 
the proper motion analysis is about 10 years; that is, the ambient shock has 
traveled 3--5 cells. In 
Figure~\ref{fig:d40d41}, the top panels show the density at $t$ = 2.2 -- 7.3 
$\times$ 10$^{-2}$. 
The simulation begins at $t$ = 0. and the shock first hits 
the cloud at $t$ = 0.005. The radius of the cloud is $\approx$ 500 cells. 
Therefore, the ambient shock has been traveling for $\sim$ 2000 yr when it 
reaches the cloud midpoint. 

From the X-ray derived shock velocity of $\sim$ 
300 km s$^{-1}$, \citet{patnaude02} estimated the age of the interaction to 
be $\sim$ 1200 years, so the modeled cloud size and the shock velocity appear 
reasonable. At the current epoch, the forward shock is 1$\arcmin$ -- 2$\arcmin$ ahead 
of the cloud shock. From Figure~\ref{fig:d40d41}, the cloud shock lags behind 
the blast wave by 10\% (bottom, Fig.~\ref{fig:d40d41}, Panel {\it b}). This 
corresponds to a physical distance of 1.9 $\times$ 10$^{17}$ cm, or 
$0 \farcm 5$ at a distance of 550 pc. By Panel {\it c} of Model 5, the blast 
wave is 20\% farther along than the cloud shock. Here, the morphology of Model 
5 closely matches that of the southwest cloud (Fig.~\ref{fig:regions}).

The observed internal cloud structure in the H$\alpha$ image
(Fig.~\ref{fig:regions}) is not that unlike the modeled shock
cloud internal structure seen in Figure~\ref{fig:d40d41}. In general, the
cloud-shocks seen in the H$\alpha$ image are $\lesssim$ $ 0 \farcm 5 $ tip to
tip. This scale is consistent with the approximate size of the internal shocks
seen in our Model 5 (Fig.~\ref{fig:d40d41}). The cloud shocks have survived the 10 years
between observations. The models, however, show that internal shocks are
straightened out over a course of a few hundred time steps ($\sim$ 200 yr).
However, over the short time we are concerned with here, the shock structure of the
cloud shock looks remarkably similar to that of the southwest cloud.



\section{Conclusions}
\label{sect:3conc}

A relatively isolated, low-density ISM cloud situated along the southwest limb
of the Cygnus Loop provides a particularly clear view of the early stages of a
SNR shock -- ISM cloud interaction.  The combination of multi-epoch
observations and high resolution numerical modeling of this cloud has provided
some new insights regarding how shocks overrun ISM clouds.  The southwest cloud's
isolation and low-density has also allowed us to view its internal density
structure and make inferences concerning the cloud's initial density
distribution. 

Our specific findings are:

1) Using multi-epoch H$\alpha$ observations of a small, isolated ISM cloud in
the southwest portion of the Cygnus Loop, we measured proper motions of
Balmer-dominated shock filaments which wrap around the cloud, as well as the
proper motion of several internal cloud shocks.  The Balmer-dominated filaments
have transverse velocities of $\sim$ 200--250 km s$^{-1}$, while the shock
filaments internal to the cloud have transverse velocities of 65 -- 140 km
s$^{-1}$. 

2) The shocked cloud's morphology does not show many of the dynamical
instabilities predicted by previous shock-cloud models.  This suggests
that there is no abrupt boundary or edge for diffuse ISM clouds. A sharp
density rise between the cloud and the ISM would lead to steep velocity
gradients at the shocked cloud -- shocked ISM interface. These steep gradients
would in turn lead to the onset of Kelvin-Helmholtz instabilities, which are
not observed. This conclusion contrasts with the shock-cloud interaction seen
in the southeast of the Cygnus Loop, where the blast wave is thought to be 
interacting with a large, dense cloud, and instability growth is clearly seen
along the cloud-shock boundary.

3) Our model hydrodynamic simulations suggest that ISM clouds are best modeled
as a constant or smoothly varying core density embedded in lower density
envelope which tapers to the surrounding ISM.  Ram pressure equilibrium
arguments suggest a cloud--ISM density contrast for this cloud of 
$\chi$ = 5 -- 17, with the
lower $\chi$'s corresponding to the diffuse regions of the cloud and the upper
limit of 17 corresponding to the dense cloud clumps.

4) A definite spacing of dense, small ``cloudlets'' inside the cloud is needed
to generate the cloud's internal morphology as seen in the H$\alpha$ image.  As
pointed out by \citet{poludnenko02}, clumps spaced too closely together interact
with the shock as if they were one large clump, while those spaced too far
apart behave as a set of individual clouds. Our models are consistent 
with the optimal
spacing of $d_{crit}$ $\approx$ 4$a_{cloudlet}$ \citep{poludnenko02}. 
The observed internal cloud
shock diffraction caused by these cloudlets is a short lived phenomena. As the
cloud shock interacts with the cloudlets, the diffracted shocks re-order
themselves on a time scale of order a few cloudlet crossing times.

\acknowledgements
We wish to thank John Blondin for both making the VH-1 code available, and
answering several questions regarding its use, and 
John Raymond for useful suggestions regarding our results. We also 
thank the anonymous referee for several helpful comments during the 
preparation of this manuscript.

\clearpage

\begin{deluxetable*}{crcrcrcr} 
\tablecolumns{8} 
\tablewidth{0pc} 
\tablecaption{Measured Proper Motions and Estimated Shock Velocities} 
\tablehead{ 
\multicolumn{4}{c}{Non-radiative Filaments} & \multicolumn{4}{c}{Cloud Filaments} \\
\cline{1-4} \cline{5-8} \\
\colhead{Filament} & \colhead{$\Delta_r$\tablenotemark{a}} & \colhead{$\mu$} & \colhead{$V_s$\tablenotemark{b}} &
\colhead{Filament} & \colhead{$\Delta_r$\tablenotemark{a}} & \colhead{$\mu$} & \colhead{$V_s$\tablenotemark{b}} \\
\colhead{Region} & \colhead{(mas)} & \colhead{(mas yr$^{-1}$)} & \colhead{(km s$^{-1}$)} & \colhead{Region} & \colhead{(mas)} & \colhead{(mas yr$^{-1}$)} & \colhead{(km s$^{-1}$)}}
\startdata
F1   & 630 $\pm$ 55 & 55 $\pm$ 5 & 140 $\pm$ 10 & C1 & 620 $\pm$ 170 & 55 $\pm$ 15 & 140 $\pm$ 30 \\
F2   & 610 $\pm$ 55 & 55 $\pm$ 4 & 140 $\pm$ 10 & C2 & 325 $\pm$ 55 & 30 $\pm$ 3 & 80 $\pm$ 10 \\
F3   & 745 $\pm$ 55 & 65 $\pm$ 3 & 170 $\pm$ 10 & C3 & 575 $\pm$ 55 & 50 $\pm$ 4 & 130 $\pm$ 10 \\
F4   & 665 $\pm$ 55 & 60 $\pm$ 3 & 155 $\pm$ 10 & C4 & 410 $\pm$ 55 & 35 $\pm$ 2 & 90 $\pm$ 10 \\
F5   & 810 $\pm$ 55 & 70 $\pm$ 5 & 180 $\pm$ 10 & C5 & 400 $\pm$ 55 & 35 $\pm$ 5 & 90 $\pm$ 10 \\
F6   & 910 $\pm$ 80 & 80 $\pm$ 7 & 210 $\pm$ 15 & C6 & 500 $\pm$ 55 & 45 $\pm$ 5 & 120 $\pm$ 10 \\
F7   & 1075 $\pm$ 110 & 95 $\pm$ 9 & 250 $\pm$ 20 & C7 & 340 $\pm$ 85 & 30 $\pm$ 6 & 80 $\pm$ 15 \\
F8   & 960 $\pm$ 80 & 85 $\pm$ 6 & 220 $\pm$ 15 & C8 & 550 $\pm$ 85 & 50 $\pm$ 7 & 130 $\pm$ 15 \\
F9   & 780 $\pm$ 55 & 70 $\pm$ 2 & 180 $\pm$ 10 & C9 & 380 $\pm$ 160 & 35 $\pm$ 13 & 90 $\pm$ 30 \\
F10  & 1025 $\pm$ 55 & 90 $\pm$ 7 & 235 $\pm$ 10 & C10 & 425 $\pm$ 130 & 40 $\pm$ 11 & 105 $\pm$ 25 \\
F11  & 1010 $\pm$ 110 & 90 $\pm$ 10 & 235 $\pm$ 20 & C11 & 280 $\pm$ 85 & 25 $\pm$ 8 & 65 $\pm$ 15 \\
F12  & 900 $\pm$ 140  & 80 $\pm$ 12 & 210 $\pm$ 25 & C12 & 295 $\pm$ 55  & 25 $\pm$ 2 & 65 $\pm$ 10 \\
F13  & 1110 $\pm$ 55 & 100 $\pm$ 4 & 260 $\pm$ 10 & C13 & 380 $\pm$ 55 & 35 $\pm$ 4 & 90 $\pm$ 10 \\
F14  & 1120 $\pm$ 85 & 100 $\pm$ 6 & 260 $\pm$ 15 & C14 & 410 $\pm$ 110 & 40 $\pm$ 10 & 105 $\pm$ 20 \\
F15  & 745 $\pm$ 55 & 65 $\pm$ 4 & 170 $\pm$ 10 & & & & \\
F16  & 625 $\pm$ 55 & 55 $\pm$ 4 & 145 $\pm$ 10 & & & & \\
F17  & 800 $\pm$ 55 & 70 $\pm$ 3 & 185 $\pm$ 10 & & & & \\
F18  & 715 $\pm$ 55 & 60 $\pm$ 2 & 155 $\pm$ 10 & & & & \\
F19  & 690 $\pm$ 55 & 60 $\pm$ 4 & 155 $\pm$ 10 & & & & \\
F20  & 720 $\pm$ 110 & 60 $\pm$ 9 & 155 $\pm$ 20 & & & & \\
F21  & 660 $\pm$ 55  & 60 $\pm$ 4 & 155 $\pm$ 10 & & & & \\
\enddata
\tablenotetext{a}{1992.6 -- 2003.7}
\tablenotetext{b}{Shock velocities assume a distance of 550 pc \citep{blair05}}
\label{tab:cc}
\end{deluxetable*} 

\clearpage

\begin{deluxetable*}{lcccccc}
\tablecolumns{7} 
\tablewidth{0pc} 
\tablecaption{Model Parameters} 
\tablehead{
\colhead{Model} & \colhead{$a_0$} & \colhead{$\chi_{max}$} & \colhead{$M$} &
\colhead{$n_{cloudlets}$} & \colhead{$\chi_{cloudlets}$} & \colhead{$a_{cloudlets}$}}
\startdata
1 & 0.30 & 6  & 20 & \nodata & \nodata & \nodata \\
2 & 0.25 & 10 & 10 & \nodata & \nodata & \nodata \\
3 & 0.30 & 6  & 20 & 13      & 10      & 0.03    \\
4 & 0.30 & 3  & 20 & 13      & 10      & 0.03    \\
5 & 0.35 & 10 & 10 & 15      & 10      & 0.05    \\
6 & 0.35 & 8  & 10 & 15      & 15      & 0.03    \\
\enddata
\tablenotetext{a}{Models 1, 3, and 4 have cylindrical density distributions.
Models 2, 5 and 6 have distributions corresponding to 
Equation~\ref{eq:densdist}.}
\label{tab:modparms}
\end{deluxetable*}

\clearpage

\clearpage

\clearpage

\clearpage

\clearpage

\begin{thebibliography}{} 

\bibitem[Blair et al.(1999)]{blair99} Blair, W.\ P., Sankrit, R., Raymond, J.\ C., 
\& Long, K.\ S.\ 1999, \aj, 118, 942 

\bibitem[Blair et al.(2005)]{blair05} Blair, W.\ P., Sankrit, R., \& Raymond, J.\ C.\ 1999, \aj, 129, 2268 

\bibitem[Blondin \& Knerr(1992)]{blondin92} Blondin, J.~M., \& 
Knerr, J.~M.\ 1992, Bulletin of the American Astronomical Society, 24, 1227 

\bibitem[Colella, \& Woodward(1984)]{colella84} Colella, P., \& Woodward, P.\ R.\
1984, J.\ Comp.\ Phys., 54, 174

\bibitem[DeNoyer(1975)]{denoyer75} DeNoyer, L.\ K.\ 1975, \apj, 
196, 479

\bibitem[Fesen, Blair, \& Kirshner(1982)]{fesen82} Fesen, R.\ 
A., Blair, W.\ P., \& Kirshner, R.\ P.\ 1982, \apj, 262, 171 
\apjs, 118, 541 

\bibitem[Fragile et al.(2004)]{fragile04} Fragile, P.~C., Murray, 
S.~D., Anninos, P., \& van Breugel, W.\ 2004, \apj, 604, 74 

\bibitem[Fragile et al.(2005)]{fragile05} Fragile, P.~C., 
Anninos, P., Gustafson, K., \& Murray, S.~D.\ 2005, \apj, 619, 327 

\bibitem[Kahn(1976)]{kahn76} Kahn, F.~D.\ 1976, \aap, 50, 145 

\bibitem[Klein, McKee, \& Colella(1994)]{klein94} Klein, R.~I., 
McKee, C.~F., \& Colella, P.\ 1994, \apj, 420, 213

\bibitem[Levenson \& Graham(2001)]{levenson01}
Levenson, N.~A. \& Graham, J.~R. 2001, \apj, 559, 948

\bibitem[Mac Low et al.(1994)]{maclow94} Mac Low, M., McKee, 
C.~F., Klein, R.~I., Stone, J.~M., \& Norman, M.~L.\ 1994, \apj, 433, 757 

\bibitem[Mellema et al.(2002)]{mellema02} Mellema, G., Kurk, 
J.~D., Rottgering, H.~J.~A.\ 2002, \aap, 395, L13


\bibitem[Nittman, Falle, \& Gaskell(1982)]{nittman82} Nittman,
J., Falle, S., P.~H.\ 1982, \mnras, 201, 833

\bibitem[Parker(1967)]{parker67} Parker, R.\ A.\ R.\ 1967, \apj, 
149, 363 

\bibitem[Patnaude et al.(2002)]{patnaude02} Patnaude, D.~J., 
Fesen, R.~A., Raymond, J.~C., Levenson, N.~A., Graham, J.~R., \& Wallace, 
D.~J.\ 2002, \aj, 124, 2118 

\bibitem[Poludnenko, Frank, \& Blackman(2002)]{poludnenko02} 
Poludnenko, A.~Y., Frank, A., \& Blackman, E.~G.\ 2002, \apj, 576, 832 

\bibitem[Stevens et al.(1992)]{stevens92} Stevens, I.~R., 
Blondin, J.~M., \& Pollock, A.~M.~T.\ 1992, \apj, 386, 265 

\bibitem[Stone \& Norman(1992)]{stone92} Stone, J.~M., \& 
Norman, M.~L.\ 1992, \apjl, 390, L17 

\bibitem[Tonry \& Davis(1979)]{tonry79}
Tonry, J. and Davis, M.\ 1979,  \aj, 84, 1511

\bibitem[Thorstensen, Fesen, \& van den Bergh(2001)]{thorstensen01} 
Thorstensen, J.~R., Fesen, R.~A., \& van den Bergh, S.\ 2001, \aj, 122, 297

\bibitem[Woodward(1976)]{woodward76} Woodward, P.~R.\ 1976, \apj, 
207, 484

\end{thebibliography}
\end{document}